# Magnon-mediated interlayer coupling in an all-antiferromagnetic junction

Yongjian Zhou[1,3], Liyang Liao[1,3], Xiaofeng Zhou[1], Hua Bai[1], Mingkun Zhao[2], Caihua Wan[2], Siqi Yin[1], Lin Huang[1], Tingwen Guo[1], Lei Han[1], Ruyi Chen[1], Zhiyuan Zhou[1], Xiufeng Han[2], Feng Pan[1] & Cheng Song[1,3]*

[1]Key Laboratory of Advanced Materials, School of Materials Science and Engineering, Beijing Innovation Center for Future Chips, Tsinghua University, Beijing 100084, China.

[2]Beijing National Laboratory for Condensed Matter Physics, Institute of Physics, University of Chinese Academy of Sciences, Chinese Academy of Sciences, Beijing 100190, China.

[3]These authors contributed equally to this work.

*Corresponding author. E-mail: songcheng@mail.tsinghua.edu.cn

The interlayer coupling mediated by fermions[1] in ferromagnets brings about parallel and anti-parallel magnetization orientations of two magnetic layers, resulting in the giant magnetoresistance[2,3], which forms the foundation in spintronics and accelerates the development of information technology[4]. However, the interlayer coupling mediated by another kind of quasi-particle, boson, is still lacking. Here we demonstrate such a static interlayer coupling at room temperature in an antiferromagnetic junction $Fe_2O_3/Cr_2O_3/Fe_2O_3$, where the two antiferromagnetic $Fe_2O_3$ layers are functional materials and the antiferromagnetic $Cr_2O_3$ layer serves as a spacer. The Néel vectors in the top and bottom $Fe_2O_3$ are strongly orthogonally coupled, which is bridged by a typical bosonic excitation (magnon) in the $Cr_2O_3$ spacer. Such an orthogonally coupling **exceeds the category of traditional collinear interlayer coupling via fermions in ground state, reflecting the fluctuating nature of the magnons, as supported by our magnon quantum well model. Besides the fundamental significance on the quasi-particle-mediated interaction, the strong coupling in an antiferromagnetic magnon junction makes it a realistic candidate for practical antiferromagnetic spintronics and magnonics with ultrahigh-density integration[5,6].**

The particle in solid matters can mediate the coupling between two separated magnetic layers. The interlayer coupling not only reflects the nature of the particle in turn[1,7], but also brings innovations to the microelectronics industry[4]. The most well-established interlayer coupling is mediated by fermions (electrons), known as Ruderman-Kittel-Kasuya-Yosida (RKKY) interaction[1], which leads to parallel and antiparallel alignment of magnetizations in two ferromagnets (FM1 and FM2) separated by a non-magnetic spacer (Fig. 1a) and the discovery of giant magnetoresistance (GMR)[2,3,7]. In such an interlayer coupling, almost all of the electrons around Fermi level are involved, resulting in a large coupling energy (~1 meV/unit cell)[2,7]. Thus, the interlayer coupling between ferromagnets can be presented as a magnetic field ($H$) ~0.01 Tesla (T)[2,3,7] where the Zeeman energy ($M\,H$) is comparable with the coupling energy. In comparison, the interlayer coupling through another quasi-particle, boson, which possesses much lower coupling energy (at the order of 0.01 meV/unit cell)[8], is a question of interest lacking experimental insight. Unfortunately, such a coupling is unfavorable in ferromagnets with large $M$ and resultantly ultralow or even indiscernible coupling field. Recently, antiferromagnets (AFM) with subtle or no net moment have attracted growing interests. The breakthroughs, especially on the manipulation and detection of antiferromagnetic moments[9–13], enable us to embody a low coupling energy to a sizable interlayer coupling field and to explore the boson-mediated interlayer coupling (Fig. 1b).

However, the absence of net moment and insensitiveness to magnetic field pose a tremendous challenge on unravelling antiferromagnetic interlayer coupling. As a pioneering work, it is highly significant to use AFM with Néel vectors that are accessible to control and readout for the coupling layers, as well as an insulating spacer to suppress possible fermion-mediated couplings. α-$Fe_2O_3$ is an antiferromagnet with a weak in-plane anisotropy and concomitant low spin-flop field[12,13], as well as sizable spin Hall magnetoresistance (SMR) signals to record its Néel vector[14–16]. Here, we demonstrate unprecedentedly orthogonal coupling of Néel vectors between two antiferromagnetic α-$Fe_2O_3$ layers in a $Fe_2O_3$/$Cr_2O_3$/$Fe_2O_3$ junction via magnetotransport measurements and x-ray magnetic linear dichroism (XMLD) spectra. The coupling is mediated by a typical boson, magnon[17–21], in the antiferromagnetic $Cr_2O_3$ spacer, which is supported by a magnon quantum well model. The interlayer coupling in an all-antiferromagnetic magnon junction makes it a realistic candidate for practical antiferromagnetic spintronics and magnonics, which have attracted increasing attention because of their potential applications on ultrahigh-density information storage and high frequency devices[5,10,22–25].

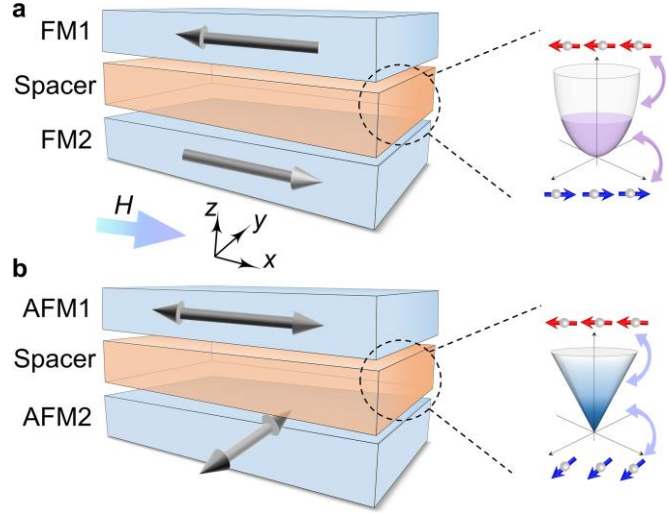

**Fig.1. Two types of interlayer coupling in magnets**. **a,** Illustration of the anti-parallel interlayer coupling mediated by fermion in ferromagnetic trilayers. The magnetization in the top and bottom ferromagnetic layers (FM1 and FM2) is anti-parallel for energy minimum. The inset denotes the coupling through fermion, which obeys Fermi-Dirac distribution. **b,** Illustration of the interlayer coupling between antiferromagnets (AFM1 and AFM2) mediated by a typical boson (magnon) found here. The Néel vectors in the two antiferromagnets are orthogonal. The interlayer coupling mediated by boson, which obeys Bose-Einstein distribution, is marked in the inset.

We first show in Fig. 2a SMR signals of a control sample $Fe_2O_3$(12 nm)/Pt(4 nm), where the magnetic field ($H$) and current ($I$) is along $x$-axis and the spin polarization generated by the spin Hall effect of Pt is along $y$-axis. As expected, comparatively low resistance states at high magnetic fields reflect that the Néel vector ($n$) of $Fe_2O_3$ is perpendicular to $H$ and $I$ due to the spin-flop at high fields and deviates towards the current direction at low fields, which is quite characteristic for negative SMR of AFMs[14–16]. The resistance peaks at approximately $\mu_0H = \pm0.35$ T owing to the deviation of $n$ from the spin-flop state appears at a negative field as sweeping the field from positive to negative (black line), indicating that the Néel vector almost keeps the spin-flop state at zero-field[12]. Note that $Fe_2O_3$ with the thickness below tens of nanometers maintains easy-plane anisotropy without Morin transition[12,14,16]. Similar SMR signals are obtained in another control sample $Cr_2O_3$(4.4 nm)/$Fe_2O_3$(4 nm)/Pt(4 nm) (Fig. 2b), where $Fe_2O_3$ was grown on a $Cr_2O_3$ buffer to ensure a closer scenario as the top $Fe_2O_3$ in the $Fe_2O_3$/$Cr_2O_3$/$Fe_2O_3$ junction. The SMR signals of the control samples are simulated and the hysteresis is due to the existence of strong Dzyaloshinskii-Moriya interaction (DMI) in $Fe_2O_3$[26]. The antiferromagnetic $Cr_2O_3$ buffer possesses a high spin-flop field higher than 6 T (Ref. [23]), which does not contribute to the observed SMR signals.

Figure 2c displays a representative HAADF-STEM (high-angle annular dark-field scanning transmission electron microscopy) image of the $Fe_2O_3$(12)/$Cr_2O_3$(4.4)/$Fe_2O_3$(4) (units in nanometers) cross-section, reflecting the epitaxial growth of the junction. Figure 2d presents the SMR curves of the $Fe_2O_3$/$Cr_2O_3$/$Fe_2O_3$ junction, which was covered by 4 nm-thick Pt. Four typical $H$ ((i) → (iv)) are denoted in the inset. The most eminent feature is that two resistance peaks exist when sweeping $H$ from positive to negative (black line) or reverse (red line), which is different from the SMR signals of a single $Fe_2O_3$ layer in Fig. 2a, b. A low resistance is obtained for $n \perp I$ ($n$ is parallel to spin polarization) at the spin-flop state. As $H$ is swept downward, the first resistance peak (high resistance state) at a positive $H$ ($\mu_0 H$ = +0.3 T, (i)) reveals that $n$ deviates from the spin-flop state and is unexpectedly aligned along the magnetic field and current ($n // I$). This observation indicates that another effect suppresses the magnetic field effect. We attribute the overwhelming effect to the interlayer coupling between two $Fe_2O_3$ layers through the $Cr_2O_3$ spacer. The AFM coupling generates an orthogonal (90°) arrangement of the Néel vectors in two $Fe_2O_3$ layers. Based on both magnetic field (Fig. 2a, b) and angle dependent SMR measurements in $Fe_2O_3$/Pt and $Cr_2O_3$/$Fe_2O_3$/Pt control samples, we find the top $Fe_2O_3$ possesses a lower spin-flop field than its bottom counterpart, in analogy to a soft ferromagnet with a small coercivity. Therefore, the Néel vector in the top $Fe_2O_3$ has the priority to deviate from the spin-flop state at a sufficient low $H$ as a result of the interlayer coupling, resulting in the resistance peak before zero-field. This is bolstered by the simulation based on calculating the energy profile of different magnetic configurations in Fig. 2e.

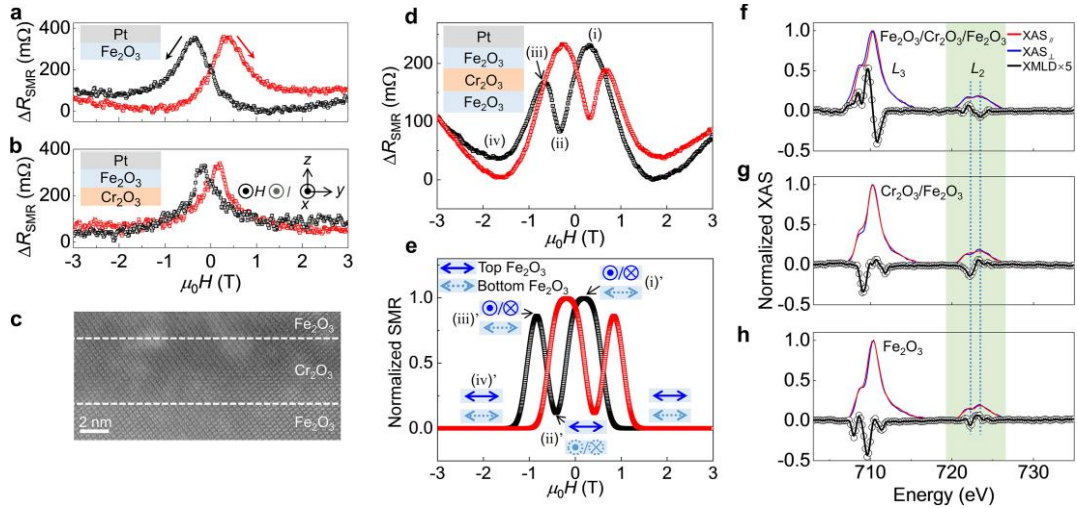

**Fig.2. SMR and XMLD results of antiferromagnetic junctions**. **a,b,** Magnetic field dependent SMR curves in control samples $Fe_2O_3$/Pt (**a**) and $Cr_2O_3$/$Fe_2O_3$/Pt (**b**) at 300 K. $\Delta R_{SMR}$ denotes the difference between resistance and the minimum. Inserts are experimental set-up. **c,** HAADF-STEM image of the $Fe_2O_3$/$Cr_2O_3$/$Fe_2O_3$ junction. **d,** SMR signals of $Fe_2O_3$/$Cr_2O_3$/$Fe_2O_3$/Pt samples at 300 K. **e,** Simulated SMR curve of $Fe_2O_3$/$Cr_2O_3$/$Fe_2O_3$/Pt samples at 300 K. Inserts are diagram of magnetic configurations at typical magnetic fields. **f–h,** Normalized XAS and XMLD spectra of $Fe_2O_3$/$Cr_2O_3$/$Fe_2O_3$ (**f**), $Cr_2O_3$/$Fe_2O_3$ (**g**) and $Fe_2O_3$ (**h**) samples. The XMLD spectra were taken from the differences of XAS spectra ($XAS_\perp - XAS_{//}$) and then multiply by a factor of 5 at the absorption edges for clarity. The highlighted region denotes Fe-$L_2$

edge and the vertical dotted lines are guide for eyes to mark the valley and peak in XMLD curves.

As $H$ sweeps to the negative side, the SMR signal decreases and a resistance valley appears at negative $H$ (ii), which is almost the same as the location of resistance peak in Fig. 2a. The direction of $n$ in the bottom $Fe_2O_3$ is $n // H$, and the interlayer coupling drives the Néel vector in the top $Fe_2O_3$ to $n \perp H$ (*I*), again giving rise to the orthogonal configuration between the two $Fe_2O_3$ layers ((ii)' in Fig. 2e). In this case, the spin current is reflected at the interface between Pt/top $Fe_2O_3$, leading to the relatively low resistance. Then $n$ in the bottom $Fe_2O_3$ rotates towards the spin-flop state ($n \perp H$) due to the increasing of negative $H$, and $n$ in the top $Fe_2O_3$ deviates towards $n // H$ (*I*) due to the interlayer coupling ((iii)' in Fig. 2e), resulting in the absorption of spin current and the second resistance peak (iii). It should be clarified that the second peak can exist when the following condition is satisfied: The coupling energy is large enough to overcome the Zeeman energy of the top $Fe_2O_3$ at the valley (ii), otherwise the Néel vector in the top $Fe_2O_3$ will maintain spin-flop state rather than deviating towards $n // H$. The magnitude of the second peak is smaller than the first one can be ascribed to less component of $n$ along $x$-axis. Since α-$Fe_2O_3$ has three-fold easy-axes[13,27], the orthogonal interlayer coupling is intrinsically a situation where the Néel vector along $x$-axis (0°) in one $Fe_2O_3$ layer, while along two other easy axes (60° and 120°, respectively) in the coupled layer with the main projection along $y$-axis (90°) for simplicity.

Besides magneto-transport measurements, we further confirm the AFM interlayer coupling by direct Néel vector characterizations. Fe *L*-edge XMLD spectra were used to detect the Néel vector of the top $Fe_2O_3$ (several nanometers-thick sensitivity) in the $Fe_2O_3/Cr_2O_3/Fe_2O_3$ junction, where 2 nm-thick Pt was deposited on top. The XMLD spectra were recorded at zero-field after applying a high magnetic field along the *x*-axis for the sake of the non-volatile feature of the Néel vector in easy-plane $Fe_2O_3$[12]. X-ray was vertically incident to the film and the polarized direction of the x-ray was parallel to the film plane. XMLD signals are obtained as $XMLD = XAS_\perp - XAS_{//}$, where $XAS_{//}$ and $XAS_\perp$ denote the x-ray absorption spectroscopy (XAS) recorded with polarization parallel with *x*-axis (//) and *y*-axis ($\perp$), respectively. Corresponding data are presented in Fig. 2f, where $L_2$-edge is highlighted because it is generally used for the Fe-based XMLD spectra analysis[28,29]. Remarkably, Fe $L_2$-edge XMLD spectrum exhibits a zero–positive–negative–zero feature, which is quite a characteristic for the Néel vector along the parallel direction (*n* // *x*-axis)[28,29], rather than the direction determined by the spin-flop (*y*-axis). The Néel vector in the top $Fe_2O_3$ aligned along *H* confirms the interlayer coupling. The scenarios differ dramatically for the control samples $Fe_2O_3$ and $Cr_2O_3/Fe_2O_3$, in which identical experiments were carried out, but an opposite polarity at $L_2$-edge (Fig. 2g,h, respectively), namely zero–negative–positive–zero, was observed, suggesting that the Néel vectors in $Fe_2O_3$ are mainly aligned along the spin-flop direction (*n* // *y*-axis) without interlayer coupling. Again considering the three-fold easy-axes of $Fe_2O_3$, the Néel vectors in the control samples should be aligned along the other two easy-axes with the main components along the direction perpendicular to the magnetic field rather than along it.

We now turn towards the temperature dependence of SMR measurements in $Fe_2O_3/Cr_2O_3/Fe_2O_3/Pt$ samples. Figure 3a shows the SMR results at various temperatures. At a relatively high temperature (*T* = 270 K), there exists two resistance peaks as we discussed above for *T* = 300 K, but the intensity of the second peak is lower than that at $T = 300$ K. Such a tendency continues with further decreasing temperature to 250 K, producing a tiny peak (or just a protruding), accompanied by the absence the second peak at 200 K. Also visible is that the location of the first resistance peak shifts towards $H = 0$ with decreasing $T$ but maintains at positive $H$, reflecting $n // H$ in the top $Fe_2O_3$ before zero-field. This behavior discloses that although the interlayer coupling between the two $Fe_2O_3$ layers persists at low temperatures, the coupling energy decreases, resulting in the dominant spin-flop state and the disappearance of the second resistance peak. This phenomenon is similar with the temperature dependence of the spin fluctuation in $Cr_2O_3$ spacer[30].

In addition to the $H$-dependent SMR signals we have explored the interlayer coupling between antiferromagnets by in-plane angle($\alpha$)-dependent SMR. Corresponding data of the $Fe_2O_3/Cr_2O_3/Fe_2O_3/Pt$ sample at $T = 300$ K measured at two typical magnetic fields of 0.5 T and 1 T are shown in Fig. 3b where $\alpha = 0°$ means $H // I$. For $\mu_0 H = 1$ T, the SMR signals exhibit a negative polarity with the valley at $\alpha = 0°$, which is a typical feature for the negative SMR of antiferromagnets[14,15]. This finding discloses that the Néel vector in $Fe_2O_3$ layer is at the spin-flop state. The situation differs dramatically for $\mu_0 H = 0.5$ T, the SMR curve exhibits a positive polarity, indicating that the Néel vector of the top $Fe_2O_3$ maintains $n // H$ due to the dominant interlayer coupling. This finding coincides with the results of field dependent SMR. In contrast, the polarity of SMR keeps negative in the control sample $Cr_2O_3/Fe_2O_3/Pt$, reflecting the antiferromagnetic feature of $Fe_2O_3$ and the absence of the interlayer coupling. Identical angular dependence measurements were carried out in the $Fe_2O_3/Cr_2O_3/Fe_2O_3/Pt$ sample with $\mu_0 H = 0.5$ T at various temperatures. The polarity of SMR is positive at high temperatures ($T = 350$ and 300 K). With decreasing temperature to 250 K, the SMR signals become quite weak or even noisy, because of a competition between the interlayer coupling ($n \parallel H$) and the magnetic field-induced spin-flop ($n \perp H$). This is accompanied by the typically negative SMR induced by the spin-flop with further decreasing temperature to 230 K and 200 K. The polarity of the control sample $Cr_2O_3/Fe_2O_3/Pt$ is always negative at different temperatures, reflecting the stable antiferromagnetic SMR of $Fe_2O_3$ without the interlayer coupling.

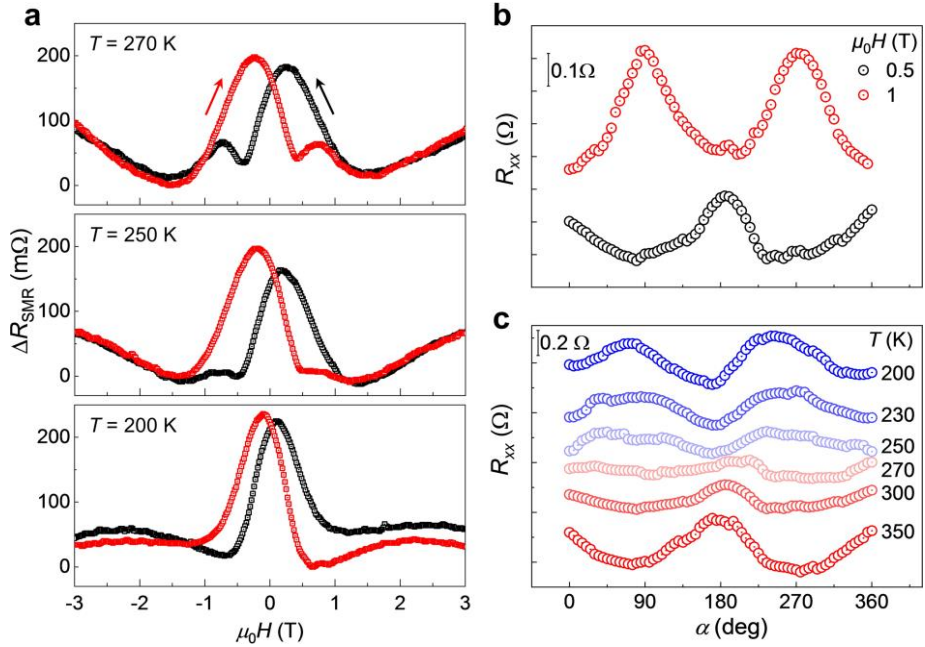

**Fig.3. Temperature dependent SMR signals. a**, SMR signals as a function of magnetic fields for the $Fe_2O_3/Cr_2O_3/Fe_2O_3/Pt$ sample at various temperatures. **b**, Angle dependence of SMR curves for the $Fe_2O_3/Cr_2O_3/Fe_2O_3/Pt$ sample at $\mu_0H = 0.5$ and 1 T. **c**, Corresponding SMR curves at various temperatures at $\mu_0H = 0.5$ T. The curves are shifted vertically for clarity.

It is significant to characterize the interlayer coupling strength. Considering that the existence of the first peak is the compromise between the interlayer coupling and spin-flop state, its location ($\mu_0H_{Coupling}$) as a function of temperature for different $Cr_2O_3$ thicknesses is summarized in Fig. 4a to reflect the coupling strength between the two $Fe_2O_3$ layers. The first peak persists at a positive field for all of the measured SMR curves, suggesting the orthogonal antiferromagnetic interlayer coupling when the $Cr_2O_3$ thickness ($t$) ranges 3–4.4 nm. It is also found that the coupling field has a similar temperature dependence for all of the samples used here. At low temperatures ($T \approx$ 10, 10, 100, and 125 K for $t$ = 3.0, 3.5, 4.1, and 4.4 nm, respectively), the first peak ($\mu_0H_{Coupling}$) almost locates at zero-field, reflecting the absence of interlayer coupling. With increasing temperature, $\mu_0H_{Coupling}$ increases and reaches the maximum just above room temperature, and then drops, which coincides with the spin fluctuation in $Cr_2O_3$ (whose bulk Néel temperature is 307 K)[30].

As the coupling field shifts towards $H = 0$ with decreasing temperature, the magnitude of second resistance peak, which is also related to the deviation of Néel vector in top $Fe_2O_3$, also changes (Fig. 3a). We then summarize in Fig. 4b the temperature dependent magnetoresistance (MR) for the $Fe_2O_3/Cr_2O_3/Fe_2O_3/Pt$ samples with various $t$. MR is related to the magnitude of the second resistance peak and is defined as MR = [$R$(second peak)–$R$(lowest)]/$R$(lowest), where the lowest is the minimum of the SMR curves. It can be seen that the MR exhibits a similar temperature dependence as the coupling field, strongly suggesting that the MR is also relevant to the interlayer coupling. As we discussed above (Fig. 2d), the second resistance peak is the result of competition between interlayer coupling and spin-flop state in the top $Fe_2O_3$ at the valley. When the temperature is low, the coupling energy is relatively small as compared with the Zeeman energy. Therefore, the Néel vector in the top $Fe_2O_3$ maintains spin-flop state ($n \perp H$), causing the vanishment of the second resistance peak. The vanishing temperature (MR = 0) is obviously higher than its counterpart for no coupling effect ($\mu_0H_{Coupling} = 0$), such as $T$ = 225 K and $T$ = 125 K for $t$ = 4.4 nm, respectively. With further increase of temperature, the coupling energy is enhanced and exceeds the Zeeman energy, resulting in the deviation of more Néel vectors towards $H$ and the resultant rapid rise of the second resistance peak (MR). As the Néel order is partially destroyed at high temperatures ($T > 340$ K for $t = 4.4$ nm), the coupling is reduced, accompanied by the decreasing of MR. An analogical situation occurs for the samples with thinner $Cr_2O_3$ ($t = 3.0$, 3.5, and 4.1 nm).

In the following we exclude the magnetic ordering in the $Cr_2O_3$ spacer involved in the interlayer coupling of $Fe_2O_3$. Firstly, If the interlayer coupling is mediated directly by the thickness independent in-plane components of the $Cr_2O_3$ moments, the Néel vectors of the top and bottom $Fe_2O_3$ layer should always be parallel due to the symmetric coupling at the top and bottom $Fe_2O_3/Cr_2O_3$ interface, irrespective of the interfacial coupling between $Fe_2O_3$ and $Cr_2O_3$ is parallel or perpendicular. Alternatively, if the interlayer coupling is mediated by the thickness dependent in-plane components of the $Cr_2O_3$ moments, e.g., a spiral spin structure, the in-plane component of Néel vector in $Cr_2O_3$ should be rotated by 90° in a very short distance. The limited DMI[26] hardly induces such a short-period spiral. Moreover, such a spiral spin structure cannot bring about the perpendicular coupling in a wide range of spacer thickness, but changeable coupling angles[31]. Thus the magnetic ordering can not generate the orthogonal interlayer coupling.

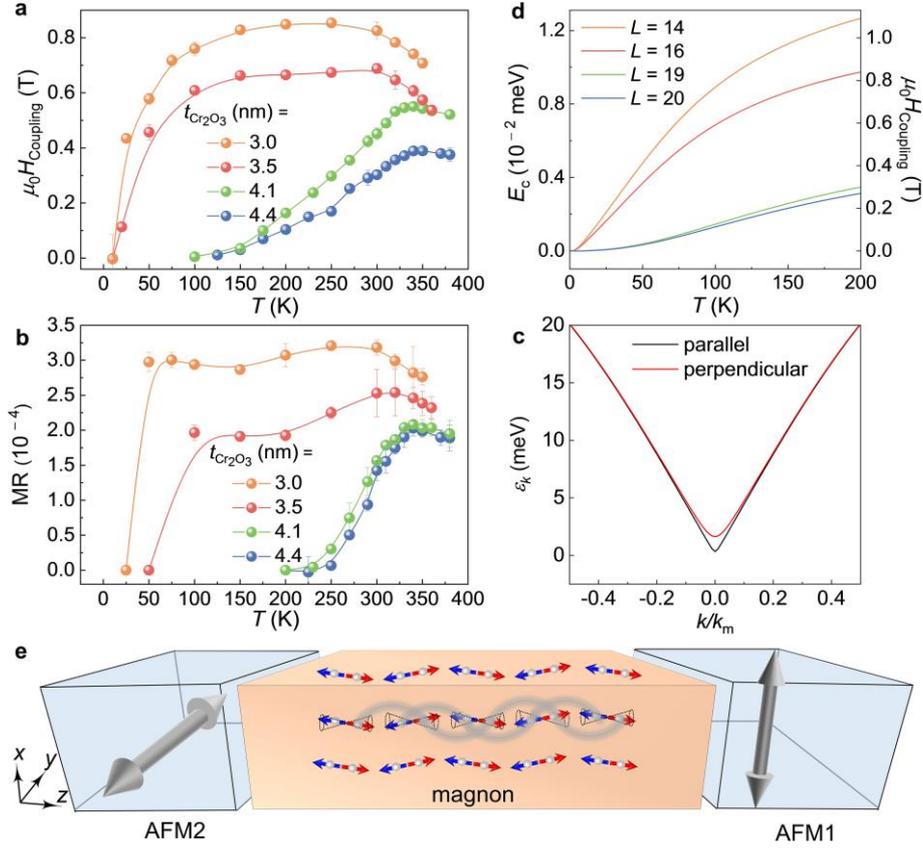

**Fig.4. Temperature and spacer thickness dependent interlayer coupling. a**, Summary of the location ($\mu_0 H_{Coupling}$) of the first peak, where $\mu_0 H_{Coupling}$ indicates the coupling strength, with various $Cr_2O_3$ layer thicknesses ($t$ = 3.0, 3.5, 4.1, and 4.4 nm). **b**, Corresponding summary of the temperature dependent magnetoresistance (MR) for $Fe_2O_3/Cr_2O_3/Fe_2O_3$/Pt samples. The error bars are estimated from the SMR data with sweeping $H$ four times. **c**, The lowest (energy) magnon band in the parallel and perpendicular states, where a representative spacer thickness of $L$ = 14 is used. **d**, Temperature dependence of the calculated coupling field for $L$ = 14, 16, 19 and 20. **e**, Schematic of the magnon-bridged orthogonal interlayer coupling in antiferromagnetic junctions.

The temperature-dependent behavior of $\mu_0 H_{Coupling}$ in Fig. 4a strongly suggests that the interlayer coupling is correlated to magnon, which is enabled by the spin fluctuation in the $Cr_2O_3$ spacer. For a magnon current in the $Cr_2O_3$ layer sandwiched by two $Fe_2O_3$ layers, the parallel and perpendicular arrangements of the Néel vectors in the two $Fe_2O_3$ layers means different boundary conditions. We use a magnon quantum well model to describe the magnon transport in the $Fe_2O_3/Cr_2O_3/Fe_2O_3$ junction. In the parallel state, the magnon current with polarization direction along the $Fe_2O_3$ Néel vector can be reflected at both $Cr_2O_3/Fe_2O_3$ interfaces, producing a quantum well mode with infinite wavelength. In contrast, in the perpendicular scenario, the magnon current with polarization direction along one $Fe_2O_3$ Néel vector can be reflected at the $Cr_2O_3/Fe_2O_3$ interface, but will be absorbed at the other interface. Thus the quantum well mode should not have infinite wavelength. Since the thermal excitation of the quantum well mode with infinite wavelength results in energy cost, the perpendicular scenario without such a mode possesses a lower energy, and becomes favorable for the interlayer coupling.

The coupling energy for the magnon quantum well model will be calculated below. The Néel vector of $Fe_2O_3$ makes the degenerated $\alpha$ and $\beta$ magnon modes in $Cr_2O_3$ recombine, forming a lowest energy mode and the other uplifted mode. The phase differences at the two $Cr_2O_3/Fe_2O_3$ interfaces of the lowest (energy) magnon mode are respectively assumed to be 0° and 90° for the parallel and perpendicular aligned $n$ in the two $Fe_2O_3$ layers. Figure 4c plots the lowest (energy) magnon band in the parallel and perpendicular states, where a representative spacer thickness of $L = 14$ is used ($L = 6$ corresponds to the $c$-axis lattice constant 1.35 nm of $Cr_2O_3$). It can be seen that the perpendicular state has a higher magnon energy $\varepsilon_k$, which would lead to a smaller contribution $\varepsilon_k[\exp(\varepsilon_k/k_BT)–1]^{-1}$ to the total energy due to the Bose-Einstein distribution, and result in the smaller total energy of the perpendicular state than that of the parallel state.

In the two samples with comparatively thin $Cr_2O_3$ ($t$ = 3.0 and 3.5 nm), the Néel vector of the $Cr_2O_3$ would have a strong tendency to rotate towards the in-plane direction due to the coupling with the $Fe_2O_3$ layers. Hence, the magnon can carry in-plane spins and form the quantum well controlled by the Néel vectors of $Fe_2O_3$. Therefore, the coupling energy starts to increase at a low temperature (~10 K). While in the two samples with thicker $Cr_2O_3$ ($t$ = 4.1 and 4.4 nm), the Néel vector of the $Cr_2O_3$ can maintain well-defined out-of-plane arrangement at low temperatures, and the magnon cannot carry in-plane spins[10,30]. Only when the thermal fluctuation is strong, can the magnon carry in-plane spins and form the magnon quantum well controlled by the Néel vectors of $Fe_2O_3$. Therefore, the coupling starts at a higher temperature. Based on these considerations, we obtain the coupling energy $E_c$ by taking the difference of the energy between the parallel and perpendicular states, as shown in Fig. 4d.

When $E_c$ is higher than the Zeeman energy, the magnon junction would be driven from the spin-flop state ($n \perp H$ for both $Fe_2O_3$) towards the perpendicular state. The Zeeman energy mainly arises from the net magnetization induced by DMI[26]. Taking the exchange field of 900 T, DMI effective field of 2 T, $L$ = 18, and 5 $\mu_B$/atom for $Fe_2O_3$[12,32,33], the coupling field is robust and estimated to be 0.86 T for $E_c$ = 0.01 meV. The temperature dependence of the calculated coupling field ($\leq$ 200 K) is also displayed in Fig. 4d, which semi-quantitatively agrees with the experimental curves. Both experimental and theoretical results disclose that a small energy difference between the parallel and perpendicular states can be embodied as sizable interlayer coupling fields for the antiferromagnetic interlayer coupling, exhibiting unique advantage as compared with its ferromagnetic counterpart. A combination of the temperature and spacer thickness dependent SMR measurements, XMLD characterizations and the magnon quantum well model demonstrates the magnon-bridged orthogonal interlayer coupling in antiferromagnetic junctions (Fig. 4e).

The discovery of the large magnon-mediated orthogonal interlayer coupling exceeds the category of traditional collinear interlayer coupling via fermions in ground state, reflecting the fluctuating nature of the magnons. The present magnon-mediated coupling in the all-antiferromagnetic junction would be a significant step towards ultrahigh-density integration of antiferromagnetic spintronics[5,6] and a new building block for magnon valves. In addition, the results disclose a possibility of manipulating the magnon band in a material through the magnetic configuration of the adjacent layers, which would help to stabilize two-dimensional magnetism[34,35]. Finally, since the bosons, such as phonon and Cooper pair widely exist in materials, our work will open a new direction for boson-mediated coupling between order parameters in condensed matter.

We thank D. Z. Hou, R. Cheng, J. Xiao, X. G. Wan, K. Shen, Y. Z. Wu, and P. Yan for fruitful discussion. We thank Beamline 08U1A of SSRF for XMLD measurements. This work was supported by the National Key R&D Program of China (Grant No. 2017YFB0405704), the National Natural Science Foundation of China (Grant No. 51871130), and the Natural Science Foundation of Beijing, China (Grant No. JQ20010).